\documentclass[letterpaper, 12pt]{article}
\usepackage{latexsym}
\usepackage{amsmath}
\usepackage{mathrsfs}
\usepackage{amssymb}
\usepackage[latin1]{inputenc}
\usepackage{graphicx}
\usepackage{color}
\def\CN{{\cal N}}

\def\Tr{{\rm Tr}\,}

\def\perp{\bot}
\def\half{\frac{1}{2}}

\def\perp{\bot}

\def\ker{{\rm ker}\,}

\newcommand{\be}{\begin{equation}}
\newcommand{\ee}{\end{equation}}
\newcommand{\bea}{\begin{eqnarray}}
\newcommand{\eea}{\end{eqnarray}}

\def\sqr#1#2{{\vcenter{\vbox{\hrule height.#2pt
    \hbox{\vrule width.#2pt height#1pt \kern#1pt
    \vrule width.#2pt}
    \hrule height.#2pt}}}}
\def\square{\mathchoice\sqr65\sqr65\sqr{2.1}3\sqr{1.5}3}

\newcommand{\bbibitem}[1]{\bibitem{#1}\marginpar{#1}}

\def\Label#1{\label{#1}%
  \smash{\hbox to0pt{\raise1ex\hbox{\tiny[#1]}\hss}}}
\def\noLabels{\let\Label=\label}
\def\nobbibitem{\let\bbibitem=\bibitem}

\title{\bf
Quiver Topology and RG Dynamics
}

\author{{\bf Vijay Balasubramanian$^{1}$, Bart{\l}omiej
Czech$^{1,2}$,} \\ {\bf Alfred D. Shapere$^{3}$, and Brian
Wecht$^{4}$ \footnote{vijay@physics.upenn.edu,
czech@sas.upenn.edu, shapere@pa.uky.edu, bwecht@ias.edu} }
\\[3mm]
\small \sl $^1$\; David Rittenhouse Laboratories, University of
Pennsylvania,
\\[-1.5mm]
\small \sl Philadelphia, PA 19104, USA \\
\small \sl $^2$\; Tata Institute of Fundamental Research,
\\[-1.5mm]
\small \sl Homi Bhabha Road, Mumbai 400 005, India \\
\small \sl $^3$\; Department of Physics and Astronomy, University
of Kentucky,
\\[-1.5mm]
\small \sl Lexington, KY 40506-0055, USA \\
\small \sl $^4$\; Institute for Advanced Study, School of Natural Sciences,
\\[-1.5mm]
\small \sl Einstein Drive, Princeton, NJ 08450, USA}

\begin{document}

\nobbibitem
\noLabels
\setlength{\baselineskip}{16pt}
\begin{titlepage}
%
%
\maketitle
\begin{picture}(0,0)(0,0)
\put(350,369){UPR-1203 -T} %
\put(350,352){TIFR/TH/08-52} %
\end{picture}
\vspace{-36pt}

\begin{abstract}
Renormalization group flows of quiver gauge theories play a
central  role in determining the low-energy properties of string vacua.
We demonstrate that useful predictions about the RG dynamics of a
quiver gauge theory may be extracted from the global structure of
its quiver diagram.  For quiver theories of a certain type,
we develop an efficient and practical
method for determining which superpotential deformations generate a flow to an interacting
 conformal fixed point.
\end{abstract}

\thispagestyle{empty}
\setcounter{page}{0}
\end{titlepage}

\tableofcontents
\newpage
\pagestyle{plain}

\section{Introduction}
\Label{sec:intro}

Product gauge theories with bifundamental matter, known as quiver
gauge theories, include many well-motivated extensions of the
Standard Model. They also arise as
effective field theories of open strings living on
stacks of $D$-branes.  The
renormalization group (RG) flows of quiver gauge theories 
from stringy scales down to
observable energies play a crucial role in determining the
low-energy physics associated with a given string vacuum.   Thus,
our ability to connect string theory usefully to low-energy
physics hinges, in part, on how well we can understand the RG
dynamics of quiver theories. Such an understanding is a
prerequisite to addressing questions like: To what extent is the
Standard Model a ``typical'' endpoint of RG flows from the
landscape of string vacua?  What TeV-scale extensions of the
Standard Model are consistent with string theory and how are they
distributed?     For example, in \cite{anarchic} it was suggested
that statistical universality in a random, but bounded, landscape
of vacua might lead to predictions for the high scale effective
field theory arising from string theory.   But to derive
predictions for the {\it low} energy world in this approach, RG
flow of generic quiver theories needs to be understood.

The problem is that even simple quiver theories may have quite
complex and unpredictable RG flows. Performing Seiberg dualities
\cite{seiberg} on different nodes of the quiver
\cite{Beasley:2001zp,seibergquiver,Fiol:2002ah,Feng:2002zw,Berenstein:2002fi,Wijnholt:2002qz}
yields many complex patterns, such as 
periodic \cite{Klebanov:2000hb} or chaotic \cite{chaoticduality} duality cascades, 
as well as structures like duality walls and trees
\cite{Franco:2003ja}. The infrared limit of a flow may depend sensitively on initial conditions, and IR fixed points may lie on conformal manifolds with many branches \cite{ls, hanany}. 
In the absence of general
techniques for precisely evaluating the flows of strongly coupled
gauge theories, it would be quite useful to find simple criteria
that could be used to infer the infrared behavior of a given
quiver theory.

The present paper attempts to extract information about the RG
dynamics of an $\CN = 1$ supersymmetric quiver gauge theory from
the global structure of its quiver diagram.  To make the problem tractable,
we choose to study equal rank quivers with $N_c$ colors and $N_f = 2N_c$
flavors at each node.   We choose these quivers 
because each node is well within the conformal window and thus,
when there is no superpotential, the theories flow to interacting
conformal fixed points in the infrared.   We give a  simple and
efficient method of determining which  superpotentials will drive
these theories to new interacting conformal fixed points and which
ones will drive unbounded flows in the original couplings.
Our results crucially involve the global structure of
the quiver and are inaccessible to the sort of node-by-node
analysis that has been the basis of most previous studies of quiver dynamics.

\section{Physical Preliminaries}
\Label{sec:physprem}
We study $\CN =1 $ equal rank $\prod SU(N_c)$
quiver gauge theories that are specified by an oriented graph $G$
with nodes $v \in V$ corresponding to gauge groups, and edges $e
\in E$ corresponding to bifundamental chiral multiplets
$\Phi^i_j$.
The theory is free from gauge anomalies if at each node the
numbers of incoming and outgoing edges are the same.    We will
only consider chiral theories since nonchiral fields will
acquire masses under generic RG flows and can be integrated out.

\subsection{Interacting superconformal fixed points}
\Label{sec:iscfp}

Our main goal in this paper will be to find a criterion for distinguishing which $\CN=1$ supersymmetric quiver theories flow to interacting superconformal fixed points in the IR.
The requirements for such a fixed point are:
\begin{enumerate}
\item All the NSVZ $\beta$-functions \cite{Novikov:1983uc} vanish.
\item All operators present in the superpotential are marginal (or irrelevant).
\end{enumerate}
When an assignment of R-charges satisfying  these conditions can
be found, we will make the  assumption that an interacting
superconformal fixed point exists. In the context of the quivers
we study in this paper, such an assumption is reasonable, although
it is not true in general that these conditions are sufficient for
the existence of a conformal fixed point. If a family of such
solutions exists, the R-charge appearing in the superconformal
algebra can be found via $a$-maximization \cite{amax}.
When a simultaneous solution to conditions 1 and 2 does not exist, we will see that
the theory is pushed toward arbitrarily strong coupling. In what follows, we will refer to the requirements implied by condition 2 as the ``marginality inequalities."

In addition to conditions 1 and 2, there is also a unitarity
requirement.  Given a gauge-invariant operator ${\mathcal O}$, superconformality requires that its
dimension $\Delta({\cal O})$ satisfies $\Delta({\mathcal O})
\geq 1$. In our theories, this implies that the R-charges of all
fields must be positive. To see this, consider the gauge-invariant
dibaryon operator \be B_{\Phi}=\epsilon_{i_1 i_2 ...
i_{N_c}}\epsilon^{j_1 j_2 ... j_{N_c}}
\Phi^{i_1}_{j_1}\Phi^{i_2}_{j_2}...\Phi^{i_{N_c}}_{j_{N_c}}. \ee
Since this operator is chiral, it satisfies $R(B_\Phi)=N_c
R(\Phi)$. In the large $N_c$ limit, $R(\Phi) < 0$ will make this
operator dramatically violate the unitarity bound. Note that
$R(\Phi)=0$ also implies a violation this bound.
Accordingly, at a superconformal fixed point it must also be true
that
\begin{equation}
R(e) > 0 \quad\forall e \in E\,.\Label{unitarityineq}
\end{equation}
If a solution to conditions 1 and 2 implies violations of the unitarity bound, we regard this as a sign that new accidental symmetries have appeared in the infrared. 

A particularly convenient class of quiver gauge theories for our purpose has  nodes which all
have equal ranks $N_c$ and satisfy $N_f = 2N_c$.
When the superpotential vanishes ($W =0$) for such a quiver, it is
reasonable to suppose that its RG flow reaches a superconformal fixed point
starting from generic initial conditions in the UV. This is
because the assumption that $N_f = 2N_c$ at each node places us
firmly in the conformal window $3N_c/2 < N_f < 3N_c$
\cite{seiberg}, and while it is possible that internodal dynamics in some cases might push some gauge couplings to IR freedom, we know of no reason to believe that this happens here.  The R-charge of each
bifundamental at the $W=0$ fixed point will be \be R(e) = 1 -
\frac{N_c}{N_f} = {1/2} \, ,\ \Label{rw0}\ee which follows from
the vanishing of the NSVZ $\beta$-functions.

In this paper our approach will be to start with the theory at the $W=0$ interacting fixed point and ask
what happens when it is deformed by the addition of relevant operators.
By eq.~(\ref{rw0}), quartic and higher order operators have
$R(\mathcal{O}) \geq 2$ and are irrelevant. Therefore, in what
follows we focus attention on theories which contain cubic
operators (mass terms are ruled out by the assumption of
chirality).

Quiver theories of this kind with $W=0$ have a moduli space of
vacua parameterized by VEVs of the matter fields that solve the
D-term equations.  At a generic point in this moduli space the
gauge symmetry is broken.   For simplicity, we will focus
attention on the origin in moduli space where the VEVs vanish and
the gauge symmetry is unbroken.

\subsection{Flows and $\beta$-functions}

All flows in the theory are driven by the $\beta$-functions. The
flow of the gauge couplings is described by the NSVZ
$\beta$-function \cite{Novikov:1983uc}
\begin{equation}
\beta_{1/g^2} = \frac{3T(G) - \sum_a T(r_a) ( 1- \gamma_a(g)) }{8 \pi^2 - g^2 T(G)},
\end{equation}
where $\Tr T^A_{r_a} T^B_{r_a} = \half T(r_a) \delta^{AB}$, the
sum is over all matter fields $\Phi_a$ in the theory, and
$\gamma_a$ is the anomalous dimension of  $\Phi_a$.  For equal-rank quivers with $N_f = 2 N_c$ at each node,
 the NSVZ beta function for the gauge coupling at node $v$ is
\begin{equation}
\beta_{1/g_v^2} =
\frac{N_c}{8\pi^2 -
g_v^2N_c} \left (1 + \half \sum_{e\sim v} \gamma_e(\{g_i\}) \right
),\Label{defbeta0}
\end{equation}
where $e \sim v$ denotes all edges entering or exiting the node.
We use the normalization $T(SU(N_c)) = N_c$ and $T(\square) =
\half$.

In the vicinity of a superconformal fixed point we can use
\begin{equation}
1 + \frac{1}{2}\gamma_e = \Delta(e) = \frac{3}{2}R(e) \Label{gammar}
\end{equation}
to express the beta functions in terms of $R(e)$.
Slightly away from a superconformal point, the R-charges on the right hand side can be defined according to the
formalism of \cite{kutasov}.
Using (\ref{gammar}), one can then write
\begin{equation}
\beta_{1/g^2_v} \propto \sum_{e\sim v} R(e) -2\, \Label{defbetag}
\end{equation}
with a positive coefficient of proportionality.  Although the coefficient includes a potentially dangerous denominator, we will assume that  we stay far away from poles in the NSVZ $\beta$-function.

Similarly, given a superpotential term $\int d^2 \theta\,
\lambda_{\mathcal O} {\mathcal O}$, the beta function for the
coupling $\lambda_{\mathcal O}$ is
\begin{equation}
\beta_{\lambda_\mathcal{O}} \propto \sum_{e \in \mathcal{O}} R(e)
- 2 \equiv R({\mathcal O}) - 2\ , \Label{betalambda}
\end{equation}
with a positive coefficient of proportionality.
Irrelevant operators are those with $R({\mathcal O}) >2$.

\section{Examples of RG Dynamics}
\Label{sec:localex}

Before investigating arbitrary $N_f = 2N_c$ quivers we will study several examples of theories in this class.
The examples are chosen to illustrate cases when a suitable choice
of superpotential forces a superconformal interacting theory to
flow to strong coupling. The main result of this
paper, derived in Sec.~\ref{sec:globalstr}, is a general characterization of superpotentials which have this
effect.

\subsection{The octahedral theory}
\Label{sec:octahedral}

Consider the vertices of the octahedron as nodes of a quiver
diagram, with edges corresponding to bifundamentals. There exists exactly one
assignment of arrows (up to charge conjugation) which
ensures that each of the eight faces of the octahedron forms a
closed loop, and thus is an eligible superpotential term. We call
this theory the octahedral
theory; see Fig.~\ref{octahedron}. Note that the outer triangle is also a face of the
octahedron and is a closed loop appearing in the
superpotential.

\begin{figure}[h]
\begin{center}
\includegraphics[scale=0.5]{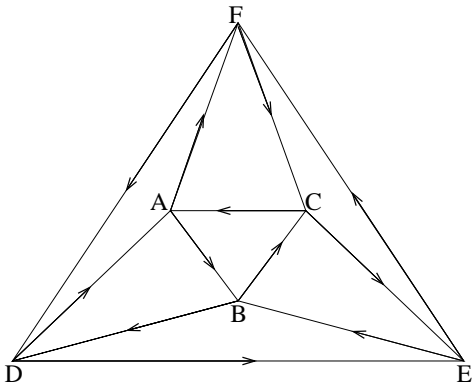}
\caption{The octahedral quiver.}\Label{octahedron}
\end{center}
\end{figure}

It is easy to see that the faces of the octahedron cannot all
simultaneously be marginal or irrelevant if the theory is
conformal. This is because the conformality condition on the gauge
coupling of the node $v$ reads:
\begin{equation}
\sum_{e \sim v} R(e) =2
\end{equation}
so that, summing over all vertices, we get:
\begin{equation}
\sum_{e \in E} R(e) = \frac{1}{2} \sum_{v \in V} \sum_{e \sim v}
R(e) = \frac{1}{2} \sum_{v \in V} 2 = 6\,. \Label{sumv}
\end{equation}
The pre-factor $1/2$ accounts for the fact that a sum over nodes
counts every bifundamental exactly twice. Likewise, if we assume
that each face is marginal or irrelevant, we can re-do the same
calculation face by face and arrive at:
\begin{equation}
\sum_{e \in E} R(e) = \frac{1}{2} \sum_{f \in F} \sum_{e \sim
f} R(e) \geq \frac{1}{2} \sum_{f \in F} 2 = 8\,, \Label{sumf}
\end{equation}
where the index $f$ runs over the faces of the octagon (cubic
terms in the superpotential) and the factor of $1/2$ again
accounts for the fact that each bifundamental participates in
exactly two loops. Because $6 \not\geq 8$, we
see that no superconformal fixed point accommodating all the cubic
superpotential terms can exist.

To see which superpotentials are compatible with an interacting
superconformal fixed point, start with
\begin{equation}
W_{\rm trial} = \lambda_1\, \mathcal{O}_{\rm ABC}\, ,
\end{equation}
where we choose to label operators (loops in the quiver) by the
nodes on which they rest. $\mathcal{O}_{\rm ABC}$ will drive the
flow to the unique R-charge assignment where it is marginal and
the NSVZ $\beta$-functions vanish:
\begin{eqnarray}
R(e_{\rm AB}) = R(e_{\rm BC}) = R(e_{\rm CA}) = R(e_{\rm DE}) =
R(e_{\rm EF}) = R(e_{\rm FD}) & = & 2/3\nonumber
\\
R(e_{\rm AF}) = R(e_{\rm FC}) = R(e_{\rm CE}) = R(e_{\rm EB}) =
R(e_{\rm BD}) = R(e_{\rm DA}) & = & 1/3\, .\Label{octsolution}
\end{eqnarray}
At this point $\mathcal{O}_{\rm DEF}$ becomes marginal, too.
However, all other cubic operators are relevant and an addition of
any one of them to the superpotential will automatically produce
an unending RG trajectory. Any pair
\begin{equation}
\{\mathcal{O}_{\rm ABC},\, \mathcal{O}_{{\rm any}\,\triangle
\neq\,{\rm DEF}}\} \Label{forbidden}
\end{equation}
is therefore forbidden from appearing simultaneously in the
superpotential if one is interested in an interacting
superconformal field theory whose low energy degrees of freedom
are described by the quiver of Fig.~\ref{octahedron}.

\begin{figure}[h!]
\begin{center}
\includegraphics[scale=0.8]{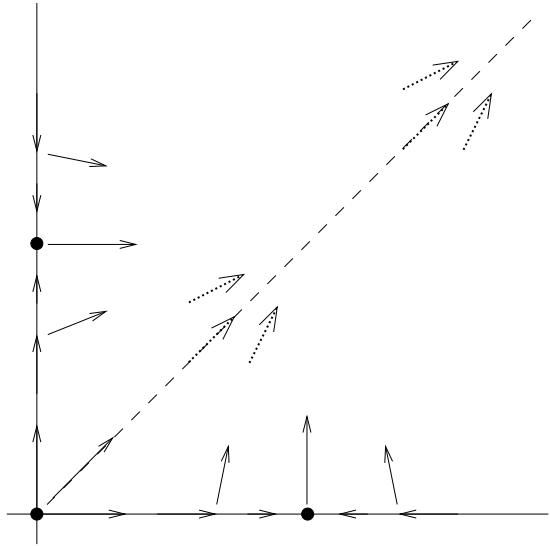}
\caption{The flow diagram of the octahedral theory drawn in the
space parameterized by two superpotential couplings, which cannot
both be turned on at an interacting superconformal fixed
point.}\Label{flowdiag}
\end{center}
\end{figure}
To understand the physics better, in Fig.~\ref{flowdiag} we draw a
flow diagram in the superpotential coupling space parameterized by
$(\lambda_1)^2,\,(\lambda_2)^2$, with $\lambda_2$ corresponding
to, say, $\mathcal{O}_{\rm ABD}$. The argument of the previous
paragraph proves that there is an interacting superconformal fixed
point when $W=\lambda_1\, \mathcal{O}_{\rm ABC}$, although it is
not stable to $\mathcal{O}_{\rm ABD}$ deformations. This is
captured by the arrows pointing along and away from one of the
axes. Because of the symmetry of the quiver, an identical set of
arrows must be drawn on the other axis, too. The $W=0$ fixed point
with $R(e)\equiv 1/2$ is of course unstable in both directions in
superpotential coupling space, as is represented by the arrows
pointing away from the origin. Overall, the flow arrows marked
with continuous lines can be deduced rigorously from the NSVZ and superpotential
$\beta$-functions.

The remaining flow vectors, shown as dotted arrows, are expected to point
away from the axes. We have already argued that there are no fixed points away from the axes, and we assume that
limit cycles in the flow do not occur (although recently discovered 
counterexamples to the $a$-theorem leave open the possibility that limit cycles could exist in four dimensions \cite{Shapere:2008zf,Shapere:2008un}).   Thus the flow must carry the
theory to infinity. 

The more complete treatment of Sec.~\ref{sec:globalstr} will
reveal that in such a situation, we expect that at least one gauge
coupling will run to infinity as well. This means that the IR
degrees of freedom are best analyzed by Seiberg dualizing
\cite{seiberg} the strongest gauge group. In the octahedral
example, dualizing any node \cite{Beasley:2001zp} gives the quiver
of Fig.~\ref{octConf}.
\begin{figure}[h]
\begin{center}
\includegraphics[scale=0.5]{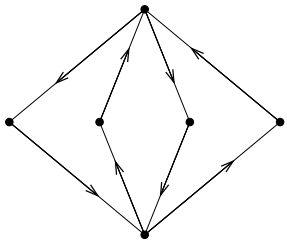}
\caption{The octahedral theory after one Seiberg
duality.}\Label{octConf}
\end{center}
\end{figure}
The four nodes on the horizontal axis of the diagram have $N_f =
N_c$ and are expected to confine, producing the same diagram as occurs in
the conifold theory \cite{kw}. The confinement, which becomes apparent only
in a Seiberg dual picture, explains why there could not exist a
fixed point whose low energy degrees of freedom would be those of
Fig.~\ref{octahedron}.

In general, we will call collections of superpotential terms like eq.~(\ref{forbidden}) which push the theory to strong
coupling {\bf forbidden sets}. A set of
terms is forbidden if the terms cannot all be made marginal or irrelevant without making the NSVZ $\beta$-functions nonvanishing.
When a forbidden set is added to the superpotential, the theory
\emph{appears to} flow without end to a strong coupling regime. We
take this to be a sign of new dynamics, which may become apparent
in, e.g., a different Seiberg duality frame.
In the remainder of this paper we shall refer to
flows resulting from forbidden sets using words like
``indefinite flow." This language is
meant as a shorthand for a situation where the degrees of freedom
characterizing an infrared fixed point are different from those
represented by the original quiver diagram.

\subsection{A locally characterized (nearly) forbidden set}

In Sec.~\ref{sec:octahedral} the forbidden sets of the octahedral
theory (\ref{forbidden}) were identified after finding the global
solution (\ref{octsolution}). In
this subsection, as an example of what we are trying to achieve, we will try to identify a forbidden set 
using local considerations alone.

\begin{figure}[h!]
\begin{center}
\includegraphics[scale=0.7]{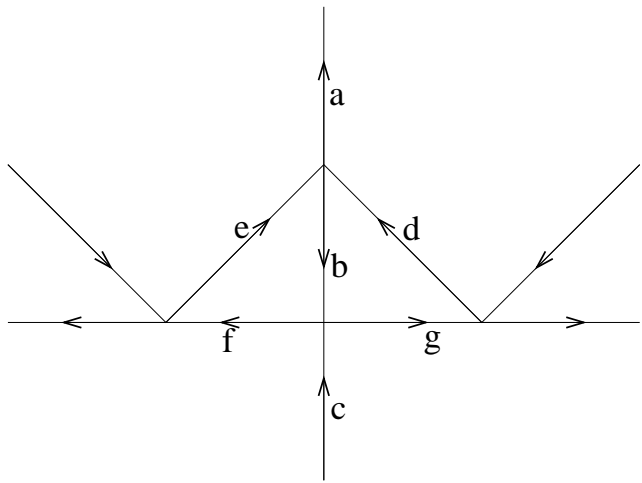}
\caption{A local obstruction to a consistent R-charge
assignment.}\Label{inconsistent}
\end{center}
\end{figure}

Consider a fragment of an $N_f = 2N_c$ equal rank quiver as shown in
Fig.~\ref{inconsistent}, and assume that the two triangular loops
are present in the superpotential. Let $a, \ldots, g$ label the
R-charges of the bifundamental fields.
If this quiver and superpotential are to be superconformal, the R-charges must satisfy
\begin{equation}
\begin{array}{rrcl}
{\rm I:} & a+ b+ d+e & = & 2\\
{\rm II:} & b+c+f+g  & = & 2\\
{\rm III:} & b+d+g & \geq & 2\\
{\rm IV:} & b+e+f & \geq & 2\, . \end{array} \Label{relations}
\end{equation}
Equations I and II set the NSVZ $\beta$-functions to zero while
III and IV ensure that the two triangular loops are marginal or
irrelevant. Now notice that the linear combination I+II-III-IV
reads:
\begin{equation}
a + c \leq 0 \, ,
\end{equation}
threatening a violation of the unitarity bound.    Indeed, the best we can do is to take $a = c =0$ in which case the dibaryons built from these two fields have vanishing dimension and hence violate the unitarity bound. Thus, the two triangles of Fig.~\ref{inconsistent} already threaten to form a forbidden set
regardless of what happens in the rest of the quiver.  Note that
this locally characterized (nearly) forbidden set is present in the
octahedral quiver, Fig.~\ref{octahedron}.  More complex examples
below will show that forbidden sets cannot always be fully characterized
by local data in the quiver.

\subsection{There are other forbidden sets}

So far we have only seen one example of a forbidden set, namely
the setup of Fig.~\ref{inconsistent}. For another example,
consider the quiver in Fig.~\ref{largerex}. Assume that all gauge
invariant terms are present in the superpotential with non-zero
couplings.

\begin{figure}[h]
\begin{center}
\includegraphics[scale=0.55]{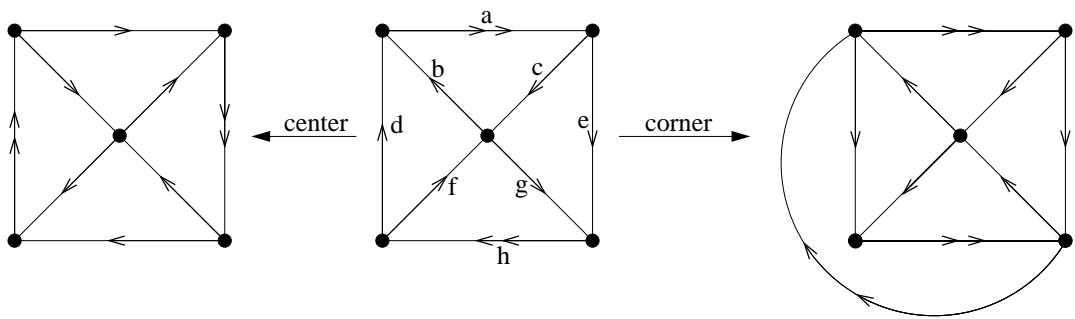}
\caption{A na\"\i vely superconformal quiver may be Seiberg dual
to one that is IR free. On the right Seiberg duality was performed
on the bottom left node.}\Label{largerex}
\end{center}
\end{figure}

The R-charges must satisfy conditions
analogous to eqs.~(\ref{relations}):
\begin{equation}
\begin{array}{rrcl}
{\rm I:} & 2a+b+d & = & 2\\
{\rm II:} & 2a+c+e  & = & 2\\
{\rm III:} & b+c+f+g & = & 2\\
{\rm IV:} & 2h+d+f & = & 2\\
{\rm V:} & 2h+e+g & = & 2\\
{\rm VI:} & a+b+c & \geq & 2\\
{\rm VII:} & f+g+h & \geq & 2\\
{\rm VIII:} & a+d+e+h & \geq & 2\,.
\end{array} \Label{relations2}
\end{equation}
The
linear combination
\begin{equation}
\frac{1}{2}({\rm I}+{\rm II}+{\rm III}+{\rm IV}+{\rm V})-({\rm
VI}+{\rm VII}+{\rm VIII}):\, 0 \leq -1
\end{equation}
reveals that it is impossible to satisfy these conditions
simultaneously.

In this case, the operators corresponding to eqs.~VI-VIII form a
forbidden set. If these terms are simultaneously present in the
superpotential, their couplings will grow without end because the
operators will never reach marginality. This is another example of
an indefinite flow like the dashed line in
Fig.~\ref{flowdiag}. In Sec.~\ref{sec:globalstr} we shall see that
in such a situation at least some subset of the gauge couplings
will grow large, too. Therefore, in the infrared a Seiberg dual
description of the theory will apply. The Seiberg dual may allow
us to identify the new dynamics of which the apparently unending RG
trajectory is a sign.

We can either dualize the central node or a corner node. But Seiberg duality on
the central node gives back the same quiver, so the only interesting nodes are the corners.
The resulting quiver is shown on the right of
Fig.~\ref{largerex}. In the dual theory, two nodes get $N_f = 3N_c$, which
suggests that the new dynamics
could be IR free. However, it may also be possible that the other nodes pull the na\"{i}vely IR free node back to being interacting, as happens with some string-derived theories.

\subsection{A global example}
\Label{sec:buildingblocks}

In most cases when a forbidden set is added to the superpotential
and the theory is pushed to strong coupling, selecting the correct
Seiberg duality frame in which to study the infrared phase is
difficult. We illustrate the difficulty of the general problem by
presenting a  formidable instance of it. Specifically, we
consider a family of quivers (defined below), which is
parameterized by an integer $b$. The family is a subset of
the class of quivers considered in this paper. Every quiver in the
family contains a forbidden set so with a suitable choice of
superpotential, each theory is driven to a strong coupling regime.
But for all $b
> 1$, the methods used in the previous examples are insufficient
for picking the right Seiberg dual picture and the tools developed
in Sec.~\ref{sec:globalstr} are necessary. As a bonus, we shall
have more to learn from the limit $b \rightarrow \infty$.

We construct our family of examples by assembling building blocks
represented in Fig.~\ref{largequiv}. Each building block is
structured around a central triangle, with three batches of lines
parallel to the three sides forming an array of surrounding
quadrangles. The building blocks may be indexed by the numbers of
parallel lines in each batch, so the block in the figure would be
denoted $\triangle_{2,\,3,\,4}$. Each triangle (quadrangle) in the
building block forms a closed loop in the quiver and therefore
represents a possible cubic (quartic) term in the superpotential.

\begin{figure}[h]
\begin{center}
\includegraphics[scale=0.25]{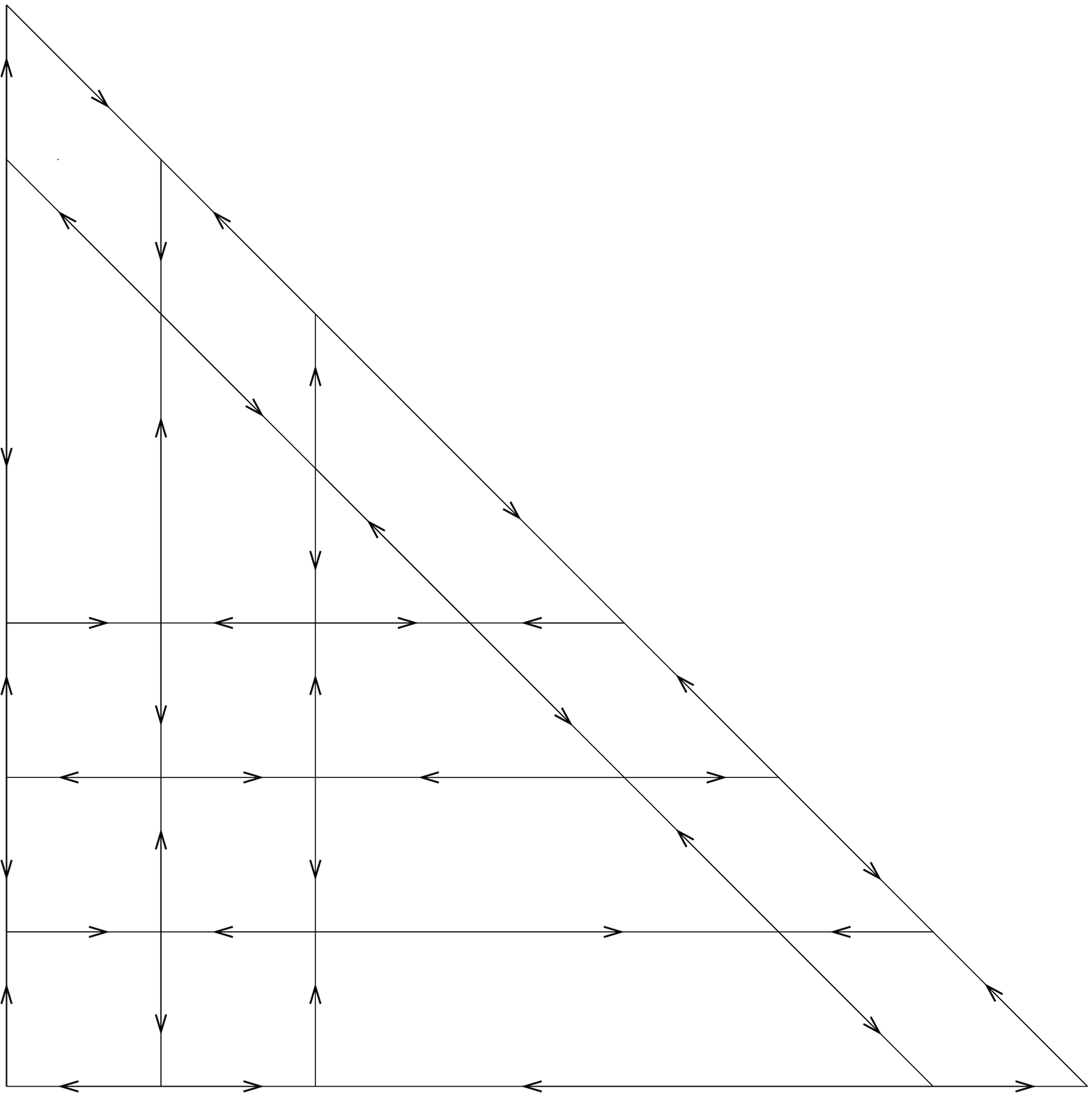}
\caption{A triangular building block
$\triangle_{2,\,3,\,4}$.}\Label{largequiv}
\end{center}
\end{figure}

Consider the octahedral theory of Fig.~\ref{octahedron}. It is
made up of eight building blocks $\triangle_{1,\,1,\,1}$ assembled
together to form an octahedron. If we now replace each building
block with $\triangle_{b,\,b,\,b}$, we shall form another quiver
diagram with $N_f = 2N_c$ on all nodes. Call the resulting quiver
$G_b$, so the octahedral diagram is $G_1$. It is easy to see that
if all the cubic and quartic terms are turned on in the
superpotential, the couplings of $G_b$ run indefinitely. This may
be seen using the fact that each $G_b$ can be embedded on a
sphere, and therefore
\begin{equation}
|F| = |E|-|V|+\chi = (2|V|)-|V| + \chi = |V|+\chi\,,
\end{equation}
where $|E| = 2|V|$ holds because exactly two edges end at each
vertex. If all the cubic and quartic operators are turned on, then
at a superconformal fixed point the following relations, analogous
to eqs.~(\ref{sumv}-\ref{sumf}), must hold:
\begin{equation}
|V| = \frac{1}{2} \sum_{v \in V} 2 = \frac{1}{2}\sum_{v \in V}
\sum_{e \sim v} R(e) = \sum_{e \in E} R(e) = \frac{1}{2}\sum_{f
\in F} \sum_{e \sim f} R(e) \geq \frac{1}{2}\sum_{f \in F} 2 = |F|
= |V|+\chi\,.
\end{equation}
Because we are on a sphere, $\chi = 2$ and such a fixed point
cannot exist. Therefore, a Seiberg dual description is again
necessary to study the theory in the IR, but now, in contrast to
the previous examples, there is no tractable method to reveal on
which node to dualize. The techniques of Sec.~\ref{sec:globalstr}
will select a likely candidate.

Importantly, in the present example the indefinite flow of the
theory is a direct consequence of the spherical structure of the
quiver, $\chi \not\leq 0$. Indeed, in $G_{b \rightarrow \infty}$
we can make arbitrarily large local patches of the quiver
perfectly superconformal, that is, imposing that the NSVZ $\beta$-functions vanish and that
all terms in the superpotential are marginal or irrelevant. We
may only conclude that the RG flow of the theory with a generic
superpotential will never rest after accounting for the fact that
the building blocks glue up together to form a sphere. The present
example illustrates explicitly that global considerations are a
necessary ingredient in a general study of renormalization group
flows.

\section{Superconformality  and Quiver Structure}
\Label{sec:graphprem}

We will show that necessary conditions for the existence of
superconformal fixed points of our gauge theories are encoded in
the oriented and unoriented incidence matrices of the quiver.

The unoriented incidence matrix $B$, which keeps track of which edges connect to which vertices,
 will turn out to be related to changes in the NSVZ
$\beta$-functions at each node. This occurs because the
$\beta$-functions do not discriminate between fundamentals and
anti-fundamentals (their quadratic Casimirs are the same), so all
one needs to know is which edges hit which nodes, and not whether
the edges are entering or exiting. We will show that the kernel of $B$ ($\ker B$) encodes R-charge assignments  (vectors in $\mathbb{R}^E$) which leave the NSVZ $\beta$-functions unchanged.

In contrast, the oriented incidence matrix $A$, which includes the
direction of connectivity via plus and minus signs, turns out to
encode information about superpotential deformations. Because
gauge-invariant operators are closed loops, the orientation of the
edges is essential for determining which operators are allowed. As
we will describe, the kernel of $A$ ($\ker A$) encodes those
R-charge assignments which are correlated with the running of
superpotential couplings.

Our objective is to scan the space $\mathbb{R}^E$ of candidate
R-charge assignments for configurations satisfying the superconformality conditions
in Sec.~\ref{sec:physprem}.
Each possible R-charge assignment is represented as a column vector
$\vec{R}^{\rm tot} \in \mathbb{R}^E$. In the class of theories we
study, the NSVZ $\beta$-functions vanish when
\be
\vec{R}^{\rm tot} = \vec{r} \equiv \frac{1}{2}\cdot\vec{1},
\Label{rdef}
\ee
where $\vec{1}\equiv (1,1,\ldots,1)$,
i.e., by  assigning
R-charge $1/2$ to each edge. We then parameterize the possible
R-charges at an interacting superconformal fixed point by
deviations from $\vec{r}$, taking
\be
\vec{R}^{\rm
tot} = \vec{r} + \vec{R} \,.
\Label{splitR}
\ee
 The role of $\vec{R}$ will be to ensure that the 
marginality inequalities and the unitarity bound are satisfied while
keeping the NSVZ $\beta$-functions zero, which is already accomplished by $\vec{r}$.

\subsection{Vanishing of the NSVZ $\beta$-functions}
\Label{vanishingbeta}

The NSVZ $\beta$-function \cite{Novikov:1983uc},
\begin{equation}
\beta_{1/g^2} = \frac{3T(G) - \sum_a T(r_a) ( 1- \gamma_a(g)) }{8
\pi^2 - g^2 T(G)}\,, \Label{betaagain}
\end{equation}
is necessarily zero at any superconformal fixed point. The numerator
of the NSVZ $\beta$-function can be usefully recast as the anomaly
in the R-current.    Thus, satisyfing $\beta_{NSVZ} = 0$ is
tantamount to having an R-anomaly free theory; this transpires
because the trace of the stress tensor is in the same supermultiplet as
the divergence of the R-current.    Because of this, any anomaly
free global $U(1)$ can be added to a candidate R-symmetry while
still satisfying $\beta_{NSVZ} = 0$. Consequently, there will be a
vector space of possible solutions to $\beta_{NSVZ} = 0$. In the
quiver gauge theories we are considering, there is one global
$U(1)$ for each of the $|E|$ edges in the quiver, but anomaly
freedom requires that the net added $U(1)$ charge vanish at each of
the $|V|$ nodes.   This gives $|V|$ linear constraints on $|E| =
2|V|$ variables for our $N_f = 2 N_c$ quivers. If the constraints
are non-degenerate, this gives a solution space of anomaly-free
R-charge assignments of dimension $|V|$. Degeneracies between the
anomaly constraints may increase the dimension of the solution
space.   In this section we will describe a simple way of
representing this solution space for our quiver gauge theories.

Recall from eq.~(\ref{defbetag}) that for $N_f = 2N_c$ quivers,
the NSVZ $\beta$-function for the gauge coupling $g_v$ is given
(up to an unimportant positive coefficient) by
\begin{equation}
\beta_{g_v} \propto 2- \sum_{e\sim v} R^{\rm tot}(e)\, .\Label{defbetagv}
\end{equation}
To represent this succintly,
define the
$|V|\times |E|$  incidence matrix $B$ by
\begin{equation}
B_{v e} = \left\{ \begin{array}{ll}
                      1 & \textrm{if edge $e \in E$ is incident to
                      vertex $v \in V$} \\
                      0 & \textrm{otherwise}\end{array} \right .
\end{equation}
Then, splitting the total R-charge as in (\ref{splitR}), the
vanishing of the $|V|$ NSVZ $\beta$-functions is written as
\begin{equation}
\beta_{g_v} \propto 2 - \sum_{e} B_{ve}\, (r^e + R^e) = 0
~~~~\Longrightarrow~~~~ B \vec{R} = 0\, ,\Label{betabr}
\end{equation}
where we used $B_{ve} r^{e} = 2$.   In other words,
\be \vec{R}
\in {\rm ker}\, B \, . \ee
Equivalently, adding $\vec{R}$ to a
candidate R-charge assigment leaves the NSVZ $\beta$-functions
unchanged because  $\vec{R}$ is the charge vector of an
anomaly-free flavor symmetry $U(1)_F$. The orthogonal complement
\be ({\rm ker}\, B)^\perp = {\rm Im}\, B^{\rm T} \ee
 comprises
R-charge assignments which necessarily make the NSVZ
$\beta$-functions nonzero, hence moving away from a superconformal
fixed point.

The kernel of the unoriented incidence matrix $B$ of the quiver
thus describes a vector space of possible R-charge assignments at
a fixed point of the gauge couplings.  What is the dimension of
this vector space?     The rank of $B$ is known \cite{incid} to be
$|V|-1$ if the graph is bipartite and $|V|$ otherwise.   (Recall
that a bipartite graph is one for which the nodes can be labeled
$+$ or $-$ in such a way that edges only connect nodes of opposite
sign.)    $B$ has reduced rank for a bipartite graph since the sum
$\sum_v (-1)^v B_{ve}=0$, where $(-1)^v$ is the label of the node.
Equivalently, a quiver can have at most one degeneracy in its
R-anomaly cancellation equations, and this happens if and only if
the quiver is bipartite.

In chiral bipartite quivers, gauge invariant operators are at least
quartic in the bifundamentals and thus cannot drive a flow from
the $W=0$ fixed point with $\vec{R}^{\rm tot} = \vec{r}$ at which
our RG flows start. For this reason, we shall fix our attention on
\emph{non-bipartite} quivers in this paper, giving ${\rm rank}\,B
= |V|$ and
\begin{equation}
\dim{{\rm ker}\,B} = |V| = \dim{{\rm Im}\,B^{\rm T}}\, .
\end{equation}
Here we used the fact that $\dim{\ker B} + {\rm rank}\, B = |E|$
and $|E| = 2|V|$ because every vertex is the initial vertex of
exactly two edges. Thus, we have $|V|$ linearly independent
$U(1)_F$ charge vectors that can be added to the R-symmetry at
the superconformal point,  along with $|V|$ linearly independent
ways of altering the NSVZ $\beta$-functions. One of the latter
modes is the vector of unities $\vec{1}$.

\subsection{Marginality of the superpotential}
\Label{marginalW}

The flows we consider start by the addition of a superpotential containing relevant operators at the $W=0$ fixed point of a quiver gauge theory.  If this theory flows to a new superconformal fixed point, all operators in the superpotential will have to become marginal or irrelevant.   Thus the $\beta$-function for all the superpotential couplings should satisfy
\begin{equation}
\beta_{\lambda_\mathcal{O}} \propto \sum_{e \in \mathcal{O}} R^{\rm tot}(e)
- 2\,  \geq  0.
\Label{margineq1}
\end{equation}
To achieve this, the R-charges of operators in the superpotential will have to shift from their values at the $W=0$ fixed point where they are determined by setting $\vec{R}^{{\rm tot}} = \vec{r}$ (\ref{rdef}), i.e. by  assigning an R-charge of $1/2$ to each bifundamental.     At a new superconformal fixed point, $U(1)$ global symmetries will mix into the R-symmetry so that the R-charge becomes $\vec{R}^{{\rm tot}} = \vec{r} + \vec{R}$ as in (\ref{splitR}).     If the superpotential has net charge zero under the $U(1)$ that is mixing in, the $\beta$-functions for the superpotential couplings will not change.
We would therefore like to identify the vector space of $U(1)$ charge assignments $\vec{R}$ under which every closed loop in the quiver is neutral.   The orthogonal complement of this space will describe charge assignments that change the $\beta$-functions of superpotential couplings.

This is most easily done in terms of the  $|V| \times |E|$ oriented incidence matrix $A$:
\begin{equation}
A_{ve} = \left\{ \begin{array}{rl}
                      1  & \textrm{if $v \in V$ is the final   vertex of the edge $e \in E$}\\
                      -1 & \textrm{if $v \in V$ is the initial vertex of the edge $e \in E$}\\
                      0 & \textrm{otherwise}\end{array} \right
                      .\Label{defa}
\end{equation}
$A$ differs from $B$ in that it encodes the directions of the
arrows in the quiver (bifundamentals) with signs. Now consider  a  charge assignment that takes the form
\begin{equation}
\vec{R} = A^{\rm T} \mathbf{v}
\end{equation}
for some $\mathbf{v} \in \mathbb{R}^V$. With this assignment, the charge of a
bifundamental $e$ is given by the difference of the values of
$\mathbf{v}$ at the final vertex $f(e)$ and the initial vertex
$i(e)$.   It follows that any gauge invariant operator that forms a closed loop in the quiver will have a vanishing charge, since the charge of the operator will be  given by a telescoping sum of the charges of the individual bifundamentals.
 Thus, any R-charge assignment in the image of $A^{{\rm T}}$ (${\rm Im}\,
A^{\rm T}$) has no effect on the $\beta$-functions of superpotential couplings.

Conversely, suppose an R-charge assignment $\vec{R}$
does not alter the $\beta$-function of any closed loop in the quiver.  Then we can
explicitly construct $\mathbf{v}$ for which $\vec{R} = A^{\rm
T} \mathbf{v}$ in the obvious way: Fix $\mathbf{v}_1 = 0$ for
definiteness, and set $\mathbf{v}_{v \neq 1}$ recursively by
imposing $\mathbf{v}_{f(e)} = R(e) + \mathbf{v}_{i(e)}$ for all $e
\in E$, starting from vertex 1. This procedure could only fail if
there were at least two paths from vertex 1 to some vertex $v$
with different R-charges $R(p_1)$ and $R(p_2)$. But then appending
any path $p$ from $v$ to vertex 1 would form two different gauge
invariant operators with R-charges $R(p_1) + R(p)$ and $R(p_2) +
R(p)$, respectively. Because all closed loops in the quiver are
assumed to have charge 0 under $\vec{R}$, $R(p_1) - R(p_2)$ must
also vanish, so the procedure of finding $\mathbf{v}$ is
guaranteed to work.

We conclude that the space of charge assignments which leave the
$\beta$-functions of superpotential couplings invariant is
precisely the image of $A^{{\rm T}}$.   $U(1)$ charge assignments
in this space, known as the {\it bond space} of the graph, cannot
affect the RG running of closed loop operators, and will not play
a role in driving a relevant superpotential towards being marginal
or irrelevant.   In contrast, the orthogonal complement of the
bond space contains the $U(1)$ charge assignments that do change
the $\beta$-functions of the superpotential and can hence be used
to satisfy (\ref{margineq1}).  This complement is the kernel of
$A$ (${\rm ker}\,A$), also known as the {\it cycle space} of the
graph.

The dimensions of ${\rm Im}\, A^{\rm T}$ and ${\rm ker}\,A$ are
known. This is because $A\,A^{\rm T} = L$ is the Laplacian matrix
of the quiver graph, which is known to have exactly one zero
eigenvalue since the quiver is connected \cite{laplacian}. Thus,
\begin{equation}
\dim{{\rm Im}\, A^{\rm T}} \geq \dim{{\rm Im}\, L} = |V|-1\, .
\end{equation}
On the other hand, $A^{\rm T} \mathbf{1} = \vec{0}$, which is
simply the statement that every edge has an equal number of
initial and final vertices. Therefore
\begin{equation}
\dim{{\rm Im}\, A^{\rm T}} \leq |V|-1
\end{equation}
and an equality follows. Thus, we have exactly $|V|-1$ linearly
independent charge assignments which do \emph{not} impact the
marginality of the superpotential and $|V|+1$ orthogonal ones that
do, i.e.,
\be {\rm dim} \, {\rm Im}\,A^{{\rm T}} = |V| - 1
~~~~~{\rm and}~~~~~ {\rm dim} \, {\rm ker}\,A = |V| + 1 \, . \ee
The latter set, of course, contains $\vec{1}$.

\subsection{The space of superconformal charge assignments}
\Label{ybasis}

To check whether one of our quivers has an R-charge assignment
consistent with a superconformal fixed point, we can pursue the
following strategy:  (i) Split the total charge as $\vec{R}^{{\rm
tot}} = \vec{r} + \vec{R}$ where $\vec{r}$ makes
the NSVZ $\beta$-functions vanish at the $W=0$ fixed point; (ii)
Choose $\vec{R}$ to have no effect on the NSVZ $\beta$-functions
while driving  the superpotential to marginality.

Recall from Secs.~\ref{vanishingbeta} and \ref{marginalW}  that
the kernel of the unoriented incidence matrix of the quiver (${\rm
ker}\,B$) encodes charge assignments that leave the NSVZ
$\beta$-functions invariant, while the charge assignments in the
kernel of the oriented incidence matrix (${\rm ker}\,A$) can be
used to drive a relevant superpotential towards being marginal. Thus
we are interested in precisely the charge assignments that are
contained in the intersection
\begin{equation}
\mathcal{Y} \equiv {\rm ker}\,B \cap {\rm ker}\,A\,;~~~~~ n \equiv
\dim{\mathcal{Y}} \, .
\end{equation}

If a solution exists, both the NSVZ $\beta$-functions and the
marginality inequalities can be satisfied by picking
$\vec{R}^{{\rm tot}} = \vec{r} + \vec{y}$ where $\vec{y} \in
\mathcal{Y}$.  But the orthogonal complements of $\mathcal{Y}$
play a role in the physics, too.  Specifically, the  orthogonal
complement of $\mathcal{Y}$ in ${\rm ker}\,A$, $\mathcal{V} \equiv
\mathcal{Y}^\perp \cap {\rm ker}\, A$, comprises all the
remaining charge assignments that change the $\beta$-functions of
superpotential couplings.  This will be of use in analyzing the
likely direction of flow of theories that escape to strong
coupling like the octahedral example in Sec.~3.   For a choice of
basis $\{\vec{v}_i\}$ of $\mathcal{V}$, note that $i$ ranges
between 0 and $|V|-n$ and choose to set $\vec{v}_0 \equiv
\vec{1}$.

Similarly the orthogonal complement of $\mathcal{Y}$ in ${\rm
ker}\,B$, $\mathcal{Z} \equiv \mathcal{Y}^\perp \cap {\rm
ker}\, B$, spans the R-charge assignments that  change neither the
$\beta$-functions for the gauge couplings nor those for the
superpotential couplings.   Thus, charge vectors $\vec{z} \in
\mathcal{Z}$ play no role in showing the {\it existence} of a
superconformal fixed point.   However, the physical charge,
determined by $a$-maximization \cite{amax} at a superconformal fixed point,
may include a component in $\mathcal{Z}$. We note that the
dimension of $\mathcal{Z}$ is $|V|-n$ and that in general
$\mathcal{V} \not\!\perp\, \mathcal{Z}$.

To find a basis for $\mathcal{Y}$, recall that vectors $\vec{y} \in \mathcal{Y}$ are characterized by
\begin{equation}
A\,\vec{y} = B\,\vec{y} = \mathbf{0}\,.\Label{kerker}
\end{equation}
Eq.~(\ref{kerker}) represents $|V|$ identical sets of 2 scalar
equations each:
\begin{eqnarray}
p + q - r - s &=& 0\\
p + q + r + s &=& 0\,.\Label{kerkervertex}
\end{eqnarray}
The variables, which refer to the R-charges of the four
bifundamentals incident on a given vertex, are presented in
Fig.~\ref{yvertex} along with the solution.
\begin{figure}[h]
\begin{center}
\includegraphics[scale=0.5]{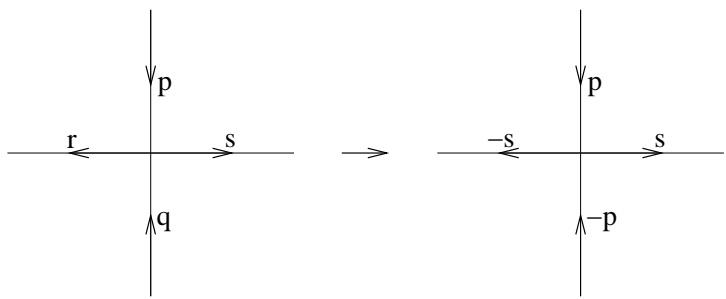}
\caption{A local solution to equations $A\,\vec{y} = B\,\vec{y} =
\mathbf{0}$.}\Label{yvertex}
\end{center}
\end{figure}
We see that on each vertex, locally, there are two independent
ways of satisfying (\ref{kerker}). One of them assigns opposite
values to the incoming arrows ($p \neq 0,\, s = 0$), and the other
assigns opposite values to outgoing arrows ($p = 0,\, s \neq 0$). Now notice that each
of the two modes can be propagated, starting from a single edge
assigned a non-zero value, to form a global solution of
(\ref{kerker}) by the following simple rule:
\begin{quote}
At each vertex, if one incoming (outgoing) arrow is assigned a
value $p$, assign the other incoming (outgoing) arrow the value
$-p$, leaving the assignments of the outgoing (incoming) arrows
unchanged.
\end{quote}
The procedure is continued until the resulting sequence of
connected edges closes back on itself, as do the dashed lines in
Figure~\ref{basiccirc}.
\begin{figure}[h]
\begin{center}
\includegraphics[scale=0.6]{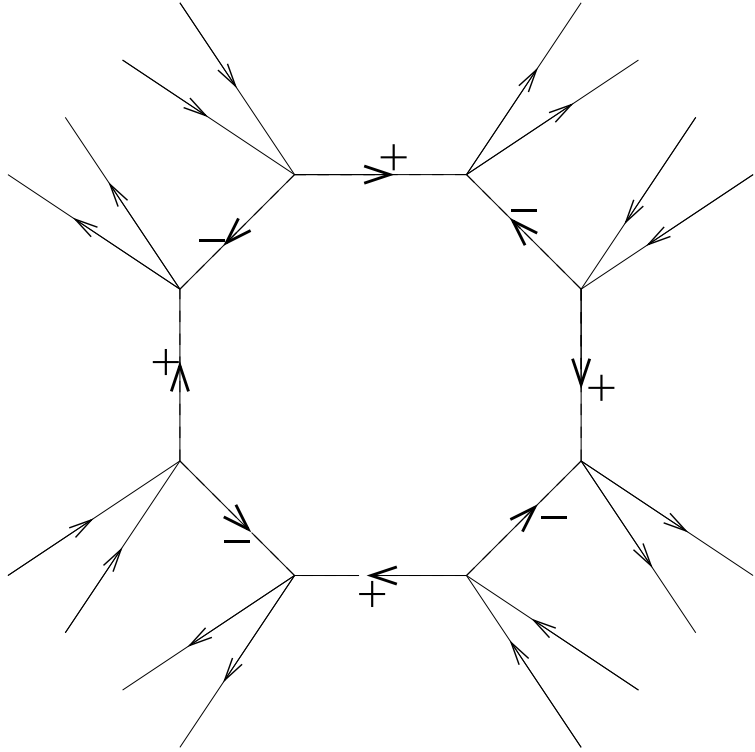}
\caption{A basic circuit $\vec{y}_i$ is marked with dashed lines.
}\Label{basiccirc}
\end{center}
\end{figure}
After one mode is constructed, one may construct another, starting
with any edge not covered by the previously constructed modes.
When no uncovered edges remain, one has obtained $n =
\dim{\mathcal{Y}}$ basis vectors $\vec{y}_i$. The modes
$\vec{y}_i$ enjoy a number of desirable properties. They are:
\begin{enumerate}
\item[-] closed lines in the quiver, consisting of alternating $+ / -$ signs,
\item[-] orthogonal\footnote{Throughout this
paper, orthogonality is defined with respect to the Euclidean
norm.} in $\mathbb{R}^E$ (by construction),
\item[-] uniquely defined,
\item[-] efficiently constructed (it takes $2|V|$ steps to identify the
vectors $\vec{y}_i$);
\item[-] they form a partition of the set of edges $E$ of the
quiver, and thus
\item[-] they form a basis of $\mathcal{Y}$.
\end{enumerate}
Motivated by these properties, we shall call the modes $\vec{y}_i$
{\bf basic circuits} of a quiver. The parameter $n$, which is the
dimension of $\mathcal{Y}$, is simply the number of basic circuits
that make up the quiver.

As an illustration of the concept of basic circuits, we re-draw
the quiver diagrams considered in Sec.~\ref{sec:localex}, marking
the basic circuits with continuous, dashed, and dotted lines. Note
that every double edge (two parallel bifundamentals) automatically
forms its own basic circuit. The building blocks of
Sec.~\ref{sec:buildingblocks} provide another convenient
illustration: we recognize the straight lines in
Fig.~\ref{largequiv} as segments of basic circuits.

\begin{figure}[h]
\begin{center}
\includegraphics[scale=0.8]{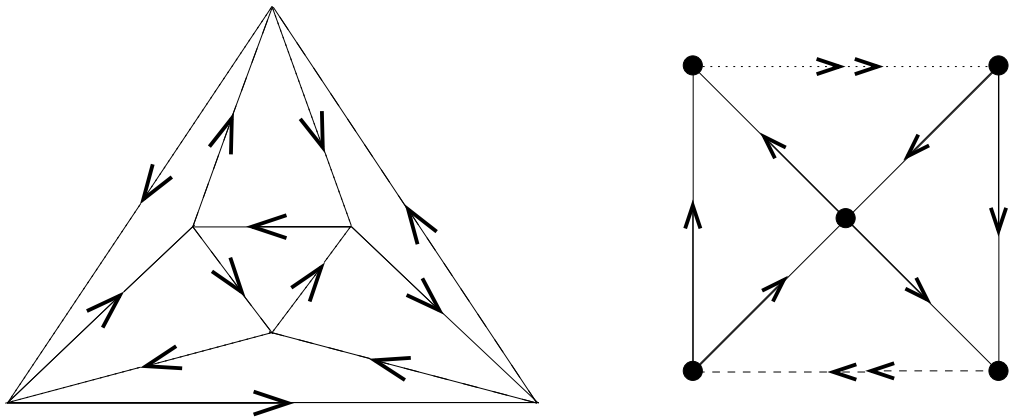}
\caption{A decomposition of the quivers considered in
Sec.~\ref{sec:localex} into basic circuits marked with continuous,
dashed, and dotted lines.}\Label{basiccircuits}
\end{center}
\end{figure}

\section{General Analysis of RG Dynamics}
\Label{sec:globalstr}

A necessary condition for the existence of an interacting
superconformal fixed point is that the superpotential does not
contain a ``forbidden set," i.e.~a set of superpotential terms
that cannot all simultaneously be marginal or irrelevant without
spoiling the conformal symmetry or unitarity. In our language,
such a forbidden set is one whose marginality inequalities are
inconsistent with the vanishing of the NSVZ $\beta$-functions and unitarity bound.

In this section we completely characterize all the forbidden sets
of an arbitrary $N_f=2N_c$ quiver theory. In most instances this
only requires analyzing the marginality inequalities and the NSVZ $\beta$-functions, since it is
generically difficult to get negative R-charges. In order to
explicitly incorporate the unitarity bound in the
formalism, it suffices to modify one equation.  We point out the
necessary modification in a footnote.

\subsection{The fixed point of a deformed theory}

In this section we will study the possible charge assignments that
preserve the NSVZ $\beta$-functions but change the superpotential
$\beta$-functions. These charges are of the form
\begin{equation}
\vec{R} = \sum_{i} c_i \vec{y}_i ~ \in ~ \mathcal{Y} \equiv \ker B
\cap \ker A, \Label{formofr}
\end{equation}
since charges in $\ker B$ do not
change the NSVZ $\beta$-functions, and charges in $\ker A$ change
marginality. These charges measure the departure from the solution
$\vec r \equiv \half  \cdot \vec 1$ of the $W=0$ theory.

Consider deforming a theory by successive additions of relevant
operators. Each time a relevant operator
$\mathcal{O}$ is added, the vector $\vec{R}$ must respond so as to
drive $\mathcal{O}$ to be marginal or irrelevant. Thus, we must
satisfy the inequality
\begin{equation}
R^{\rm tot}(\mathcal{O}) = \vec{R}^{\rm tot} \cdot
\vec{i}_\mathcal{O} = (\vec r + \vec R) \cdot \vec i_\mathcal{O} \geq 2\, . \Label{margineq}
\end{equation}
The vector $\vec{i}_\mathcal{O}$ counts how many times a given bifundamental
appears in $\mathcal{O}$. For example, for operators corresponding to minimal loops in the quiver, the
entries in $\vec{i}_\mathcal{O}$ are only 0's and 1's.

One may object that (\ref{margineq}) need not be an inequality,
but an equation, because the flow will only proceed until the
relevant operator driving the flow becomes marginal, not beyond.
However, if a theory is successively deformed by individual
operators $\mathcal{O}_1,\, \mathcal{O}_2$, it may generally
happen that the flow induced by $\mathcal{O}_2$ will drive the
operator $\mathcal{O}_1$ (which was marginal before the addition
of $\mathcal{O}_2$) to become irrelevant. To allow for this
possibility, we leave the $\geq$ sign in (\ref{margineq}).

Each operator in the superpotential introduces one marginality
inequality. A necessary condition for the existence of an interacting superconformal fixed point is that a simultaneous solution to all such inequalities exists.
One may then repeat the exercise and
deform the theory with another relevant operator. If, after adding
some number of relevant operators one by one, a solution to the resulting
inequalities fails to exist, the superpotential includes a
forbidden set and the theory will exhibit the type of run-away
behavior seen in the octahedral theory. In the remainder of this
paper we study such situations in detail.

First, we give a necessary and sufficient condition for a set of
terms to be a forbidden set. It turns out to involve the basic
circuits introduced in Sec.~\ref{ybasis}, which are defined
globally on the quiver. Second, we use the machinery of
Sec.~\ref{sec:graphprem} to study the trajectory along which the
theory escapes to strong coupling. This allows us to put forward a
natural guess about the correct Seiberg duality frame in which the
infrared phase of the theory should be studied.

\subsection{Determination of forbidden sets}
\Label{sec:determination}

To begin, it is useful to recast the marginality inequality
(\ref{margineq}) in terms of the vector $\vec{R} \in \ker B \cap
\ker A$. We will be concerned with ``excess'' dimensions of
operators, measured by how far from marginality they are. We
define
\begin{equation}
R^{\rm exc}({\cal O}) \equiv R^{\rm tot}({\cal O})-2,
\end{equation}
which can be written as
\begin{equation}
R^{\rm exc}(\mathcal{O}) = (\vec{R} + \vec{r}) \cdot
\vec{i}_\mathcal{O} - 2 = \sum_{i=1}^{n} c_i \vec{y}_i \cdot
\vec{i}_\mathcal{O} + p_\mathcal{O}. \Label{marineqy}
\end{equation}
Here we have used (\ref{formofr}) and defined $p_\mathcal{O}$ to
be the excess R-charge of the operator $\mathcal{O}$ above
marginality at the $W=0$ fixed point, where all R-charges are 1/2.
For example, $p_\mathcal{O} = -\frac{1}{2}$ for cubic operators, 0
for quartics, $\frac{1}{2}$ for quintics, etc.\footnote{To
incorporate unitarity bounds in a similar way, one
introduces an analogue of eq.~(\ref{marineqy}) that measures the
``excess'' R-charge of a bifundamental above the unitarity bound.
It reads $R^{\rm exc}(e) = \sum_{i=1}^{n} c_i \vec{y}_i \cdot
\vec{i}_e + \sum_{i=1}^{|V|-n} b_i \vec{z}_i \cdot \vec{i}_e +
p_e\ge 0$, 
with $\vec{z}_i \in (\mathcal{Y}^\perp \cap {\rm ker}\,
B)$ and $p_e = \frac{1}{2}$.} The physical content of
(\ref{marineqy}) is simply that the net deviation from being marginal
is measured by the deviation with all R-charges set equal to 1/2, plus
however much gets added via $\vec{R}$. In terms of (\ref{marineqy}),
the marginality inequalities take the simple form $R^{\rm
exc}(\mathcal{O}) \geq 0$.

If we add several operators ${\cal O}_j$, $R^{\rm
exc}(\mathcal{O}_j) \geq 0$ must hold for each ${\cal O}_j$. The
inequalities (\ref{marineqy}) for a set of operators ${\cal
O}_j,\, j=1,...,m$ can be usefully expressed in a matrix form
\begin{equation}
\left( \begin{array}{cccc} c_1 & \ldots & c_n & 1
\end{array} \right)
\left( \begin{array}{cc}
(\vec{y}_1)^{\rm T}      & 0      \\
\vdots                   & \vdots \\
(\vec{y}_{n})^{\rm T} & 0     \\
(\vec{0})^{\rm T}      & 1
\end{array} \right)
\left( \begin{array}{cccc}
\vec{i}_{\mathcal{O}_1} & \vec{i}_{\mathcal{O}_2} & \ldots & \vec{i}_{\mathcal{O}_m} \\
p_{\mathcal{O}_1}       & p_{\mathcal{O}_2}       & \ldots & p_{\mathcal{O}_m}
\end{array} \right)
\geq
\left( \begin{array}{cccc}
0 & 0 & \ldots & 0
\end{array} \right)
\, , \Label{cyio}
\end{equation}
where the inequality should hold for each component. We can rewrite this as
\begin{equation}
\left( \begin{array}{cccc} c_1 & \ldots & c_{n} & 1
\end{array} \right)
M'
\geq
\left( \begin{array}{cccc}
0 & 0 & \ldots & 0
\end{array} \right),
\Label{rexcmatrix}
\end{equation}
where $M^\prime$ is the product of the two rightmost matrices on the lefthand side of (\ref{cyio}).
The matrix $M^\prime$ is easy to compute given our choice of basis in terms of basic circuits. Each
non-trivial entry in $M'$ is given by $\vec{y}_i\cdot
\vec{i}_{\mathcal{O}_j}$, which is simply the (signed) number
of edges that the $j^{\rm th}$ operator has in common with
the $i^{\rm th}$ basic circuit.

If each inequality in (\ref{rexcmatrix}) is satisfied, so is any linear combination of these inequalities with non-negative coefficients.
The only way in which the marginality inequalities could be
inconsistent is if some suitably normalized linear combination of the basic
inequalities produces $-1 \geq 0$. To characterize this situation
it is convenient to augment the matrix $M'$ to $M$
by adding a leading column,
\begin{equation}
M = \left( \begin{array}{cc} 0 & \qquad \\ \vdots & M' \\ 0 & \\ 1
&
\end{array} \right) =
\left( \begin{array}{cc}
(\vec{y}_1)^{\rm T}      & 0      \\
\vdots & \vdots \\
(\vec{y}_{n})^{\rm T} & 0     \\
(\vec{0})^{\rm T}      & 1
\end{array} \right)
\left( \begin{array}{ccccc}
\vec{0} & \vec{i}_{\mathcal{O}_1} & \vec{i}_{\mathcal{O}_2} & \ldots &
\vec{i}_{\mathcal{O}_m} \\
1 & p_{\mathcal{O}_1} & p_{\mathcal{O}_2} & \ldots & p_{\mathcal{O}_m}
\end{array} \right)
\, .\Label{defm}
\end{equation}

Replacing $M^\prime$ with $M$ in eq.~(\ref{rexcmatrix}) yields
\begin{equation}
\left( \begin{array}{cccc} c_1 & \ldots & c_{n} & 1
\end{array} \right)
M
\geq
\left( \begin{array}{ccccc}
1 & 0 & 0 & \ldots & 0
\end{array} \right),
\Label{rexcmatrixaug}
\end{equation}
and an inconsistent linear combination of inequalities
leads to
\begin{equation}
\left( \begin{array}{cccc} c_1 & \ldots & c_{n} & 1
\end{array} \right)
M \left( \begin{array}{c} 1 \\ w_1 \\ \vdots \\
w_m \end{array} \right) = 0, \Label{weq}
\end{equation}
for some non-negative coefficients $w_j$.
When a superconformal fixed point does not exist, this equation
holds for arbitrary values of $c_i$, so
\begin{equation}
M \left( \begin{array}{c} 1 \\ w_1 \\ \vdots \\
w_m \end{array} \right) = \left( \begin{array}{c} 0 \\ 0 \\
\vdots \\ 0
\end{array} \right)\,, \qquad w_j \geq 0  \Label{wlindep}
\end{equation}
necessarily follows from the \emph{non-existence} of a
superconformal fixed point solution. This establishes that
eq.~(\ref{wlindep}) is a necessary condition a set of terms to be
forbidden. The selection of terms forming a forbidden set enters
(\ref{wlindep}) through $M$.

It is easy to see that eq.~(\ref{wlindep}) is also a sufficient
condition, if we accept that any consistent set of R-charges
implies the existence of a superconformal fixed point (which is a
mild assumption for our $N_f=2N_c$ quivers). For if a consistent
assignment of R-charges exists, then for some choice of $c_i$
eq.~(\ref{rexcmatrixaug}) has only non-negative entries and the
right hand side of (\ref{weq}) is at least 1, so
eq.~(\ref{wlindep}) cannot hold. Thus, the part of $\ker M$ that
lives in the positive orthant\footnote{The positive orthant is the
set $\{w=(w_i)\,|\,w_i\geq 0,\,i=1,\,\ldots,\,m\}$; after
\cite{dictionary}.} defines the forbidden sets of the quiver. We
again point out that the non-trivial entries of $M$ are given by
the scalar products $\vec{y}_i \, \cdot \,
\vec{i}_{\mathcal{O}_j}$, which are exceedingly easy to compute.

\paragraph{Summary} \
\\
\newline We have shown that the inequalities resulting from requiring a set of operators to be marginal or irrelevant can be nicely summed up in a coordinate-independent language. Given a quiver diagram $G$ and a set of operators
$\{\mathcal{O}_j\}$ present in the superpotential, one can determine if the set of operators is forbidden via the following decision algorithm:
\begin{enumerate}
\item Identify the basic circuits $\vec{y}_i$ of $G$ (Sec.~\ref{ybasis}).
\item Compute the matrix $M$ according to eq.~(\ref{defm}). The columns of $M$
correspond to the operators $\mathcal{O}_j$. The non-trivial entries are given
by the (signed) number of edges that the basic circuit $\vec{y}_i$ shares with
the loop corresponding to $\mathcal{O}_j$\item Find $\ker M$ and check if it intersects the positive orthant,
eq.~(\ref{wlindep}).
\item If and only if the answer to the previous step is yes, the set $\{\mathcal{O}_j\}$ is a forbidden set.
\end{enumerate}

\subsection{Escape to strong coupling}
\Label{sec:escape}

In the octahedral theory, we noted that adding a forbidden set of
operators to the superpotential drives the theory to strong
coupling. On general grounds, we expect such behavior since there
is no interacting superconformal fixed point from the standpoint
of the original degrees of freedom. When a gauge group factor gets
strongly coupled, it is natural to perform a Seiberg duality transformation on
that node and to rewrite the theory in terms of different degrees of
freedom which better describe the IR physics. In this section, we
put forward a canonical guess for which node to first dualize.  

To arrive at this guess we pursue the following logic.   Starting at the $W=0$ fixed point with the associated superconformal R-charges, we add a ``forbidden set'' of relevant deformations characterized by their  ``deficit'' of R-charge below marginality.  The sign of the $\beta$-functions is such that RG flow will drive these deformations towards marginality.   Since we cannot dynamically solve the flow equations, we estimate the direction of flow by assuming that it will act to decrease the deficit R-charge of the superpotential deformations in a direction that maximally reduces their mutual incompatibility with marginality.  Given this direction of flow for the R-charges, we can analyze the resulting direction of flow of the NSVZ $\beta$-functions.   It is natural to guess that the group with the steepest NSVZ $\beta$-function is the first to become strongly coupled and hence the one to be Seiberg dualized.   Since we cannot
actually follow the flow to strong coupling, we have no guarantee
that the group with the steepest NSVZ $\beta$-function at
$W=0$ is really the first to become strongly coupled, so
one should regard this as a first approximation.

Specifically, suppose the superpotential contains a forbidden set
$\{\mathcal{O}_j\}$. At least some subset of these operators are
relevant; for them the quantity $R^{\rm exc}(\mathcal{O}_j)$ in
eq.~(\ref{marineqy}) is negative and represents a deficit of
R-charge which must diminish along the flow. If we insist that the $\beta$-functions for the gauge couplings vanish,  the flow can only alter the
R-charges by a vector that lies in $\mathcal{Y} = \ker B \cap \ker
A$, as in (\ref{formofr}). The precise direction in the space
$\mathcal{Y}$ along which $R^{\rm exc}(\mathcal{O}_j)$ increases
towards zero (i.e. toward marginality) most steeply is given by
the gradient of $R^{\rm exc}(\mathcal{O}_j)$ with respect to the
coefficients $c_i$. But because the set $\{\mathcal{O}_j\}$ is
forbidden, using (\ref{weq}) we have
\begin{equation}
\frac{\partial}{\partial c_i} \left(\sum_j w_j R^{\rm
exc}(\mathcal{O}_j)\right) = \frac{\partial}{\partial c_i}(-1) = 0
\Label{excy}
\end{equation}
for some choice of $w_j \geq 0$. This confirms that the modes
$\vec{y}_i$
cannot
simultaneously bring all the operators in $\{\mathcal{O}_j\}$ to
marginality, in accordance with the definition of a forbidden set.
Said differently, the parenthesis in eq.~(\ref{excy}) represents
an essential deficit of R-charge which cannot be quenched by any
vector in the charge subspace $\mathcal{Y}$.  Thus a flow proceeding along a direction in ${\mathcal Y}$ will neither affect the gauge $\beta$-functions, nor will it reduce the particular linear combination of  charge deficits in (\ref{excy}).

However, we know that the $\beta$-functions for the superpotential couplings will drive a flow until this R-charge deficit is quenched.   In Sec.~\ref{marginalW} we saw that the running of the superpotential is affected by shifting R-charges by charge vectors $\vec{R}$ that are drawn from the space $\ker A$. 
Because the argument in the previous paragraph excludes the space $\mathcal{Y}
= \ker B \cap \ker A$, the charge vector $\vec{R}$ must include a shift in a
direction contained in the orthogonal complement, $\mathcal{V} =
(\mathcal{Y}^\perp \cap \ker A)$. The effect of the modes of
$\mathcal{V}$ on the marginality of operators is captured by an
analogue of eq.~(\ref{marineqy}):
\begin{equation}
R^{\rm exc}_{\cal V}(\mathcal{O}_j) =  \sum_{i=1}^{n} a_i
\vec{v}_i \cdot \vec{i}_{\mathcal{O}_j}\Label{rexcv}
\end{equation}
The deficit of R-charge represented by the parenthesis in
(\ref{excy}) will then diminish most steeply in the direction
determined by a gradient taken with respect to
the coefficients $a_i$:
\begin{equation}
\frac{\partial}{\partial a_i} \left (\sum_j w_j R^{\rm exc}_{\cal
V}(\mathcal{O}_j) \right ) =\vec{v}_i \cdot \left ( \sum_j w_j
\vec{i}_{{\cal O}_j} \right ) \propto \Delta a_i \Label{agrad}
\end{equation}
We will assume that the flow proceeds along this line of steepest descent.  Then (\ref{agrad})
represents the relative intensities by which the respective $\vec{v}_i$ modes will be turned on in the flow.

We now wish to compute how such a change of R-charges will affect
the gauge $\beta$-functions.  
Because of eq.~(\ref{betabr}), the change
in the NSVZ $\beta$-functions is straightforwardly related to the
change in $\vec R$ via
\begin{equation} \Delta \beta_v \sim -
\sum_e B_{ve}\Delta R^e. \Label{deltabv}
\end{equation}
$\Delta \vec{R}$ is easy to compute. It follows from the change in
$a_i$, eq.~(\ref{agrad}), and is given by
\begin{equation}
\Delta \vec{R} = \sum_{i} \vec{v}_i\, \Delta a_i \propto \left (
\sum_i \vec{v}_i \vec{v}_i^T \right ) \left ( \sum_j w_j
\vec{i}_{{\cal O}_j} \right ) = {\cal P}_{\cal V} \left ( \sum_j
w_j \vec{i}_{{\cal O}_j} \right )\,,\Label{deltar}
\end{equation}
where we have used the projection operation ${\cal P}_{\cal V} =
\sum_i | v_i \rangle \langle v_i |$. The appearance of the
projection operator onto $\mathcal{V}$ fixes the normalization of
$\vec{v}_i$, which was hitherto unspecified. Substituting in
(\ref{deltabv}) gives
\begin{equation}
\Delta \beta_v \propto -B\,{\cal P}_{\cal V} \left ( \sum_j w_j
\vec{i}_{{\cal O}_j} \right )\,. \Label{deltabwithpv}
\end{equation}
In the next paragraph we demonstrate that the projector
$\mathcal{P}_\mathcal{V}$ may in fact be removed from the
resulting expression, since the linear combination $ \sum_j w_j
\vec{i}_{{\cal O}_j}$ is already in $\cal V$.

First, we notice that since each vector $\vec{i}_{\mathcal{O}_j}$ corresponds to a loop in the quiver,
\begin{equation}
A\,\vec{i}_{\mathcal{O}_j} = \mathbf{0}\,,
\end{equation}
where $A$ is the oriented incidence matrix defined in
(\ref{defa}). Therefore, the expression next to
$\mathcal{P}_\mathcal{V}$ in eq.~(\ref{deltabwithpv}) lies in the
kernel of $A$. On the other hand, because the set
$\{\mathcal{O}_j\}$ is forbidden, we know from (\ref{wlindep})
that the basis vectors of $\cal Y$ are each orthogonal to $\sum_j
w_j \vec{i}_{{\cal O}_j}$. Therefore, $\sum_j w_j \vec{i}_{{\cal
O}_j}$ lies in that part of ${\rm ker}\, A$ which is orthogonal to
$\mathcal{Y}$, that is, $\mathcal{V}$. But then
$\mathcal{P}_\mathcal{V}$ is redundant in (\ref{deltabwithpv}) and
we may write
\begin{equation}
\Delta \beta_v \sim -B \left ( \sum_j w_j  \vec{i}_{{\cal O}_j} \right ) = - B_{ve} \left ( \sum_j w_j i^e_{{\cal O}_j} \right ) .\Label{deltabeta}
\end{equation}
The resulting expression is a first approximation to the change in
the NSVZ $\beta$-functions that results from including the
forbidden set $\{\mathcal{O}_j\}$ in the superpotential.
Accordingly, the most negative entry in (\ref{deltabeta}) defines
a canonical guess for the first node to Seiberg dualize.

Notice that the expression on the right hand side of
(\ref{deltabeta}) simply counts the number of edges incident on
each vertex, weighted loop-wise by the quantities $w_j$.
Therefore, it is manifestly non-positive on each vertex. It
follows that a deformation by a relevant operator initially does
not push gauge couplings toward zero \emph{if we stick to the
original degrees of freedom}. This motivates a statement which we
have made previously several times, that including a
forbidden set in the superpotential always forces some couplings
to become strong. Of course,
after following the trajectory to the deep infrared some gauge
couplings may flow to zero. The present analysis misses this
because (a) it is only an approximation, (b) it only tells you the
\emph{initial} direction of flow, and (c) it does not capture the
effects which may become apparent in terms of the infrared
(Seiberg dual) variables.

\paragraph{Caveats} Using
eq.~(\ref{deltabeta}) to identify the first node which Seiberg
dualizes is subject to a number of caveats. First, renormalization
group flows are not guaranteed to proceed according to the above
computations, since we do not know the details of the strongly
coupled physics. While the flow must necessarily head in a
direction of decreasing R-charge deficit, we are not aware of a
physical principle which would rigorously establish that it always
proceeds in the direction of steepest ascent. More importantly,
translating an R-charge perturbation (\ref{deltar}) into
$\beta$-functions (\ref{deltabeta}) is only valid if the
coefficients of proportionality relating eq.~(\ref{defbetagv}) to
the full $\beta$-function (\ref{defbeta0}) are uniform across all
the gauge couplings $g_{v \in V}$, which does not in general hold.
That said, eq.~(\ref{deltabeta}) is still useful as a rough
indicator of the direction of flow, if only because it is
extremely easy to compute.

Another subtlety is that eq.~(\ref{deltabeta}) may be used only
for {\bf minimal forbidden sets}, i.e., those for which there
exists only one linear dependence (\ref{wlindep}). This is because
RG trajectories depend on the initial conditions in coupling
space, but our formalism only takes into account which operators
are present in the superpotential, irrespective of their
couplings. The dependence of the flow on the initial conditions
must be reflected in eq.~(\ref{deltabeta}), and that can only
happen via the quantities $w_j$. When there is only one choice of
$w_j$, it signals that all run-away trajectories resulting from
the forbidden set $\{\mathcal{O}_j\}$, \emph{regardless of initial
conditions}, asymptote to a particular trajectory defined by
(\ref{deltabeta}). The dashed line of Fig.~\ref{flowdiag} was an
example of such an asymptoting RG flow. We conclude that it is
only sensible to use eq.~(\ref{deltabeta}) when the forbidden set
contained in the superpotential is minimal.

\subsection{An example}

In this subsection we present an example application of the
methods of Secs.~\ref{sec:graphprem} and \ref{sec:globalstr}.
Consider a gauge theory given by the quiver in
Fig.~\ref{examplecalc}.
\begin{figure}[h]
\begin{center}
\includegraphics[scale=0.27]{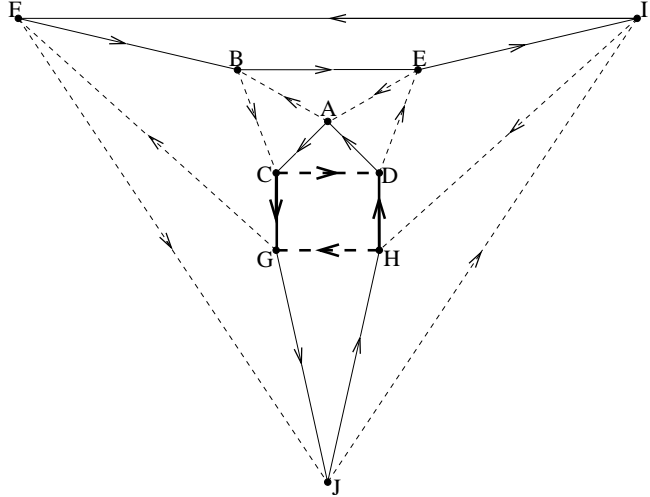}
\caption{The two basic circuits are marked with thick and thin
lines, respectively. Dashed lines carry $-$ signs and continuous
lines, $+$ signs.}\Label{examplecalc}
\end{center}
\end{figure}
It contains two basic circuits, which are drawn in thick and thin
lines, respectively. The edges charged $+1$ ($-1$) are marked with
continuous (dashed) lines.

This theory contains four cubic, seven quartic, and two quintic
gauge invariant operators, as may be verified using standard graph
theoretic techniques (see \cite{stanley} for a pedagogical
treatment). Using the technology of Sec.~\ref{sec:determination},
the matrix $M$ listing the charges of these operators under the
modes $\vec{y}_i$ is given by:
\begin{equation}
\begin{array}{rcc}
&&

\begin{array}{rrrrrrrrrrrrrr} \,\,&
\phantom{-}\!\rotatebox{-90}{\makebox[10mm][r]{\rm ABE}} &
\phantom{-}\!\rotatebox{-90}{\makebox[10mm][r]{\rm ACD}} &
\phantom{-}\!\rotatebox{-90}{\makebox[10mm][r]{\rm FJI}} &
\phantom{-}\!\rotatebox{-90}{\makebox[10mm][r]{\rm GJH}} &
\phantom{-}\!\rotatebox{-90}{\makebox[10mm][r]{\rm ABCD}} &
\phantom{-}\!\rotatebox{-90}{\makebox[10mm][r]{\rm ACDE}} &
\phantom{-}\!\rotatebox{-90}{\makebox[10mm][r]{\rm BEIF}} &
\phantom{-}\!\rotatebox{-90}{\makebox[10mm][r]{\rm FJHG}} &
\phantom{-}\!\rotatebox{-90}{\makebox[10mm][r]{\rm GJIH}} &
\phantom{-}\!\rotatebox{-90}{\makebox[10mm][r]{\rm DEIH}} &
\phantom{-}\!\rotatebox{-90}{\makebox[10mm][r]{\rm BCGF}} &
\phantom{-}\!\rotatebox{-90}{\makebox[10mm][r]{\rm ABCDE}} &
\phantom{-}\!\rotatebox{-90}{\makebox[10mm][r]{\rm FJIHG}}
\end{array}
\\[-3mm] \\
M & = & \left(
\begin{array}{rrrrrrrrrrrrrr}
0 & 0 & -1 & 0 & -1 & -1 & -1 & 0 & -1 & -1 & 1 & 1 & -1 & -1\\
0 & -1 & 2 & -1 & 2 & -1 & -1 & \phantom{-}4 & -1 & -1 & -1 & -1 & -4 & -4\\
1 & -\frac{1}{2} & -\frac{1}{2} & -\frac{1}{2} & -\frac{1}{2} & 0 & 0 & 0 & 0 & 0 & 0 & 0 & \frac{1}{2} & \frac{1}{2}\\
\end{array} \right) \end{array}
\end{equation}
We have marked the columns of $M$ with the operators to which they
correspond. The upper (middle) row of $M$ corresponds to the basic
circuit marked with thick (thin) lines in Fig.~\ref{examplecalc}.
In these rows, the $j^{\rm th}$ entry is simply the number of
continuous lines minus the number of dashed lines in a given basic
circuit that overlap with $\mathcal{O}_j$.

It is straightforward to find the non-negative coefficient linear
dependencies among the columns of $M$, as dictated by
eq.~(\ref{wlindep}). Each of them corresponds to one choice of
superpotential terms that forces the theory to run to infinite
coupling. When the coefficients $w_j$ are fed into
eq.~(\ref{deltabeta}), the result selects likely candidates for
the fastest-growing couplings and {\it eo ipso} the first nodes to
Seiberg dualize. We list the results in Table~\ref{exampleresults}
below.

\begin{table}[h]
\begin{tabular}{|l|r|l|c|l|r|l|}
\hline Forbidden set & $w_j$ & Max $-\Delta \beta_v$ & \!\!\!\! &
Forbidden set & $w_j$ & Max $-\Delta \beta_v$
\\
\hline

ABE   & 2   & B, E & \!\!\!\! & FJI   & 2   & F, I\\
BEIF  & 1/2 &      & \!\!\!\! & BEIF  & 1/2 & \\
\hline

ABE   & 1   & A, D, E & \!\!\!\! & FJI   & 1 & F, G, J\\
ACD   & 1   &         & \!\!\!\! & GJH   & 1 & \\
DEIH  & 1   &         & \!\!\!\! & BCGF  & 1 & \\
\hline

ABE   & 1   & E, H    & \!\!\!\! & FJI   & 1 & C, F\\
GJH   & 1   &         & \!\!\!\! & ACD   & 1 & \\
DEIH  & 1   &         & \!\!\!\! & BCGF  & 1 & \\
\hline

FJI   & 1   & D, I & \!\!\!\! & ABE  & 1 & B, G \\
ACD   & 1   &      & \!\!\!\! & GJH  & 1 & \\
DEIH  & 1   &      & \!\!\!\! & BCGF & 1 & \\
\hline

FJI   & 1   & H, I, J & \!\!\!\! & ABE  & 1 & A, B, C \\
GJH   & 1   &         & \!\!\!\! & ACD  & 1 & \\
DEIH  & 1   &         & \!\!\!\! & BCGF & 1 & \\
\hline
\end{tabular}
\caption{Individual boxes contain minimal forbidden sets of the
theory. Combining terms from different boxes without exhausting
any one of them does not produce a forbidden set. The remaining
columns list the choices of coefficients $w_j$ which satisfy
eq.~(\ref{wlindep}) and the nodes with the most negative value of
$\Delta \beta_v$ in eq.~(\ref{deltabeta}).} \Label{exampleresults}
\end{table}

\section{Discussion}
\Label{sec:discussion}

This paper has explored a large class of quiver theories, within
which the endpoints of renormalization group flows depend directly
on the topology of the associated quiver diagrams.  Specifically,
we found that when an $SU(N_c)^{|V|}$ quiver with $N_f=2N_c$ at all nodes
 is deformed away from the $W=0$ fixed point
by a set of relevant operators, the ensuing flow runs away to
infinite values of the couplings if and only if the operators are
chosen in such a way that (\ref{wlindep}) is satisfied for some choice
of $w_j \geq 0$.

The condition (\ref{wlindep}) depends on the basic circuits of the
quiver diagram, which are closed sequences of edges in the quiver
with alternating  signs attached to them (see Sec.~\ref{ybasis}
for a precise definition). The number $n$ of basic circuits in a
quiver is a heuristic measure of how flexible the theory is at
accommodating successive relevant deformations without losing
conformality. When $n$ is large, the theory has more freedom to
adjust its R-charges without changing the beta functions of its
gauge couplings, so as to render all relevant operators marginal.
On the other hand, when $n$ is small, a generic perturbation is
likely to produce a trajectory bound for strong coupling. Because
basic circuits form a partition of the edge set of a quiver graph,
we always have $n \geq 1$. At the other extreme, $n$ is bounded
from above by $|V|$. This bound is saturated by cyclic quiver
diagrams whose edges always appear in pairs.

It would be useful to broaden our results to a larger class of quiver theories.  For example, if we wish to fully understand the IR limits of quivers of the type we
have considered, we must also learn how to deal with the more general class of quivers
they can flow to. For example, nodes of equal-rank quivers will change rank when
the RG flow ventures onto a Higgs branch via giving VEVs to some of the
bifundamentals. Additionally, Seiberg dualizing one node
can change the number of flavors on adjacent nodes, thereby taking the quiver away from $N_f=2N_c$.

It would also be interesting to understand how the results in this
paper manifest themselves in the tiling picture of
\cite{Franco:2005rj}. For example, it seems possible that since
forbidden sets drive flows to infinite coupling, 
there might be no consistent way of dualizing a quiver
with a forbidden superpotential so as to give a tiling. This observation 
may be related to the results of \cite{Gulotta:2008ef}.

Extending our formalism to broader classes of quivers will require some thought
since the concept of a basic circuit is only
well-defined for our class. As a first step, one could consider
quivers where all nodes are $SU(N_c)$ except one which is  $SU(N_c
+1)$.  A simple quiver of this sort is the deformed conifold,
where the addition of $M$ fractional 3-branes changes the gauge
group from $SU(N)\times SU(N)$ to $SU(N+M)\times SU(N)$, and leads
to the celebrated Klebanov-Strassler
duality cascade  \cite{Klebanov:2000hb}.  
More complex quivers with unequal ranks are
believed (based on a node-by-node analysis) to admit chaotically
cascading flows \cite{chaoticduality} and other exotic RG dynamics
\cite{Franco:2003ja}.  It would be remarkable if such complex
behaviors could be deduced from the topology of the underlying
quiver diagram.

\paragraph{Acknowledgements:}
We would like to acknowledge useful conversations with Allan
Adams, Sebastian Franco, Christopher Herzog, Ken Intriligator,
David Kutasov, Hong Liu, Manavendra Mahato, John McGreevy, Costis
Papageorgakis, and David Vegh. Part of this work was done during
the Monsoon Workshop on String Theory (2008) held at TIFR. The
work of VB was supported in part by DOE grant DE-FG02-95ER40893
and in part by a Helen and Martin Chooljian membership at the
Institute for Advanced Study, Princeton. BC would like to
acknowledge the hospitality and support of the Tata Institute of
Fundamental Research under the Visiting Student program. 
AS gratefully acknowledges support from the Ambrose Monell Foundation 
and the Institute for Advanced Study, where some of this work was done.
AS is also partially supported by NSF grants 
PHY-0555444 and PHY-0245214.  BW is
supported in part by DOE grant DE-FG02-90ER40542 and the Frank and
Peggy Taplin Membership at the Institute for Advanced Study.

\end{document}